\documentclass[11pt]{article}
\usepackage{hep,epsfig}

\input{axodraw.sty}
\usepackage{axodraw}
\usepackage{cite}
\usepackage{amsmath,amssymb}
\def\be{\begin{equation}}
\def\ee{\end{equation}}
\def\bea{\begin{eqnarray}}
\def\eea{\end{eqnarray}}

\begin{document}
\vspace*{4cm} \title{FINITE MASS CORRECTIONS IN ORBIFOLD GAUGE
THEORIES~\footnote{Talk given at the XXXVIIth Rencontres de Moriond,
{\it Electroweak Interactions and Unified Theories}, Les Arcs, March 9-16,
2002}}

\author{G.~v.~GERSDORFF, N.~IRGES and M.~QUIROS}

\address{Instituto de Estructura de la Materia (CSIC), Serrano 123\\
E-28006 Madrid, Spain}

\maketitle\abstracts{The Standard Model Higgs boson can be identified
with the extra dimensional component of a gauge boson in a higher
dimensional theory where the gauge group is broken to the Standard
Model group by the orbifold action. In that case the Standard Model
symmetry can be radiatively broken by the Hosotani mechanism and the
Higgs boson mass is protected from bulk quadratic divergences by the
higher dimensional gauge theory without any need of
supersymmetry. However the latter does not protect a priori the Higgs
mass from brane quadratic divergences. We show by an explicit
calculation in the orbifold $S^1/Z_2$ that such effects are absent at
one-loop. Moreover we identify the symmetry that protects such brane
mass terms to all orders in perturbation theory and thus guarantees
the absence of a corresponding mass counterterm.}

\section{Introduction}
%\subsection{Producing the Hard Copy}\label{subsec:prod}
The mechanism of electroweak symmetry breaking (EWSB) is one of the
fundamental problems in particle physics. In particular, why this
mechanism occurs at all or why the Higgs mass is much smaller than the
{\it cutoff} of the low-energy effective theory: in fact the latter
problem is dubbed as {\it hierarchy problem}.

In the Standard Model the sensitivity of the Higgs mass on the cutoff
is quadratic, which translates into an accute hierarchy problem. In
fact one could solve the hierarchy problem by an (unnatural)
fine-tuning of the parameters of the theory, which means that the
hierarchy problem is mainly a {\it naturalness} problem. From that
point of view the hierarchy problem is alleviated in effective
theories from low-scale strings since there the cutoff is in the TeV
range~\cite{Lykken:1996fj}.  As we said, in these theories the
hierarchy problem is alleviated but not solved if there is sensitivity
of the electroweak observables to the cutoff. This is because the
ultraviolet physics is unknown from the point of view of effective
theories.

In four-dimensional theories two main solutions to the hierarchy
problem have been proposed:
\begin{itemize}
\item
A non-perturbative solution where the Higgs is a QCD-like
fermion-condensate: technicolor.
\item
A perturbative solution where an extra symmetry cancels the quadratic
divergences of the Standard Model: supersymmetry.
\end{itemize}

Supersymmetry is not the only symmetry that can protect the Higgs
mass. In the presence of compact extra
dimensions~\cite{Antoniadis:1990ew} of length $\sim 1/$TeV a {\it
gauge} symmetry can do the job leading to envisaging a perturbative
non-supersymmetric
solution~\cite{Hosotani:1983xw,Hatanaka:1999sx,Antoniadis:2000tq,Antoniadis:2001cv}
to the hierarchy problem~\footnote{Of course this statement is made
modulo the existence of a mechanism (involving the gravitational
sector) that fixes the radii of the extra dimensions, as we will
assume in this note.}. In that case the Higgses should be identified
with the (zero modes of) extra dimensional components of
higher-dimensional gauge bosons. In particular for $\delta$ extra
dimensions, $i=1,\dots,\,\delta$ there are $\delta$ bosons
\begin{equation}
\label{higgses}
H_i(x^\mu)=A_i^{(0)}(x^\mu)
\end{equation}
transforming in the adjoint representation of the higher-dimensional
gauge group. One of them should play the role of the Standard Model
Higgs. In fact if $H_i$ acquires a vacuum expectation value the
Standard Model is spontaneously broken by the Hosotani
mechanism~\cite{Hosotani:1983xw}.

\section{Electroweak Hosotani breaking}

The previously defined $H_i$ field is massless at the tree-level. This
obviously results from the higher-dimensional gauge invariance. The
masslessness of the $H_i$ field should hold at any order in
perturbation theory if the extra dimensions were infinitely
large. However for compact extra dimensions one expects that radiative
corrections generate finite~\cite{Antoniadis:1998sd} squared mass terms as $1/R^2$, where $R$ is the
radius of the extra dimension. This term goes away in the limit
$R\to\infty$ and one recovers the result of flat
dimensions. Similarly, a quadratically divergent term is forbidden for
the same reason. This phenomenon is well known in thermal field
theory~\footnote{Here the time in the Euclidean four dimensional
theory is compactified on a circle.} where the temperature plays the
role of the inverse radius.

The Hosotani breaking is finite for smooth manifolds, as toroidal
compactifications. However toroidal compactifications are problematic
and we need to compactify extra dimensions on {\it orbifolds} mainly
for two reasons~\footnote{There exist solutions to this problem
involving smooth manifolds in the presence of non-trivial
backgrounds~\cite{dvali}.}:
\begin{itemize}
\item
Higher dimensional theories are non-chiral. We need to project out the
chiral partners of the Standard Model fermions: this is performed by the
orbifold action.

\item
The Standard  Model Higgs is not in the adjoint representation of the
gauge group. We need on general grounds a gauge group $\mathcal{G}$ in
the bulk that can be broken by the orbifold action to a subgroup
$\mathcal{H}$ that coincides with, or contains, the Standard Model
gauge group and where the Higgs fields $H_i$ are $SU(2)$ doublets with
the correct hypercharge assignments.
\end{itemize}

The latter comments lead to a sort of minimal scenario that has been
worked out in Ref.~\cite{Antoniadis:2001cv} and that can be embedded
in type I string constructions. The starting point is the
higher-dimensional gauge group $\mathcal{G}=U(3)_c\times U(3)_w$. The
$SU(3)_w$ contained in $U(3)_w$ is broken down by the orbifold action
into $SU(2)\times U(1)$ while the $SU(3)_c$ in $U(3)_c$ commutes with
the orbifold action. The Standard Model Higgses are identified by the
decomposition of the adjoint representation of $SU(3)_w$ into
$SU(2)\times U(1)$ as
\begin{equation}
{\bf 8}={\bf 3}_0+{\bf 1}_0+{\bf 2}_3+{\bf\bar 2}_{-3}
\label{ocho}
\end{equation}
In this way the Standard Model gauge bosons are defined as the
zero-modes of ${\bf 3}_0^\mu+{\bf 1}_0^\mu$ while the Higgs bosons are
identified as the zero-modes of ${\bf 2}_3^i+{\bf\bar 2}_{-3}^i$ after
the orbifold breaking.

The theory on the brane contains the Standard Model gauge group and
two extra anomalous $U(1)$'s. They are broken by a generalized
Green-Schwarz mechanism that cancels the anomalies and give masses to
the corresponding $U(1)$ gauge bosons. The anomaly-free $U(1)$
coincides with the Standard Model hypercharge and is given by a linear
combination of the three original $U(1)$'s. Assuming unification of
all gauge couplings, equal to $g$, at the string scale the hypercharge
coupling is related to $g$ as $g_Y^2=3 g^2/11$ and the electroweak
angle at the string scale is
\begin{equation}
\label{angle}
\sin^2\theta_w=\frac{3}{14}\simeq 0.24
\end{equation}
This value is compatible with a multi-TeV string
scale~\cite{Antoniadis:2000en}.

So far so good. We have seen that in a Hosotani breaking the
higher-dimensional gauge invariance protects squared masses for the
Higgs sector $H_i$ from quadratic divergences in the bulk. However, as
we have seen, chirality requires compactification on an orbifold. An
orbifold has fixed points with four-dimensional boundaries where
radiative effects can be localized. Unlike radiative corrections in
the bulk, those localized on the branes are consistent with the
symmetries of a four-dimensional theory and therefore we should worry
about the possible appearance of quadratic divergences localized on
the branes for the Higgs boson masses. The symmetries of the
four-dimensional gauge theory valid on the brane do not protect Higgs
boson masses from quadratic divergences, as happens for the Standard
Model. In such a case, in spite of the protection in the bulk, we
could re-create the hierarchy problem on the branes and the
phenomenological relevance of the Hosotani breaking to solve the
hierarchy problem would be jeopardized. A detailed calculation is then
relevant. We present it in the next section where a simple $Z_2$
orbifold is analyzed in great detail.

\section{Gauge theories on orbifolds}
We have analyzed in detail brane effects on an orbifold
$\mathcal{M}_4\times S^1/Z_2$ that breaks an arbitrary gauge group
$\mathcal{G}$ to its subgroup $\mathcal{H}\subset
\mathcal{G}$~\cite{vonGersdorff:2002as}. We split the generators of
$\mathcal{G}=\{T^A\}$ into the generators of the unbroken subgroup
$\mathcal{H}=\{T^a\}$ and the broken generators of the coset
$\mathcal{K}=\mathcal{G}/\mathcal{H}=\{T^{\hat a}\}$. Our parity
assignment is defined by
\begin{eqnarray}
A_{M}^{A}(x^{\mu},-x^5)&=&\alpha^M \Lambda^{AB} A_{M}^{B}(x^{\mu},x^5)
\hskip .5cm (\rm{no\ sum\; over\; } M)
\label{Lambdaalpha}\\
%\end{equation}
%
%\begin{equation}
c^{A}(x^{\mu},-x^5)&=&\Lambda^{AB}
c^{B}(x^{\mu},x^5)
\label{pghost}\\
%\end{equation}
%
%\begin{equation} 
\Psi(x^{\mu},-x^5)&=&
\lambda_R \otimes (i\gamma^5)\Psi(x^{\mu},x^5), 
\label{pfermion}
\end{eqnarray} 
where $\Lambda$ and $\lambda_R \otimes (i\gamma^5)$ represent the
$Z_2$ action on the gauge bosons and FP-ghosts, and the fermions
respectively ($\lambda_R$ acts on the representation indices),
$\gamma^5=diag(-i,i)$ and $\alpha^{\mu}=+1$, $\alpha^{5}=-1$. In
addition, $\Lambda$ is a symmetric matrix that squares to one: it is
an orthogonal matrix. On the other hand $\lambda_R$ is a hermitian
matrix that squares to one: it is a unitary matrix.

Consistency of the 5D gauge symmetry with the orbifold action requires
the condition~\cite{Hebecker:2001jb}
\begin{equation} f^{ABC}=\Lambda^{AA'}\Lambda^{BB'}\Lambda^{CC'}f^{A'B'C'},
\label{automorphism1}
\end{equation} 
where summation over repeated indices is understood.  The above
constraint comes from the requirement that under the $Z_2$
action $F_{MN}^A\rightarrow \alpha^M\Lambda^{AB}F_{MN}^B$ (no sum over $M$), so
that $F_{MN}^AF^{AMN}$ is invariant and it is straightforward to check
that it is an automorphism of the Lie algebra of ${\cal G}$.
With no loss of generality we can diagonalize
$\Lambda^{AA'}=\eta^A\delta^{AA'}$ with $\eta^A=\pm 1$, consequently
Eq.~(\ref{automorphism1}) takes the simpler form
\begin{equation} 
f^{ABC}=\eta^{A}\eta^{B}
\eta^{C}f^{ABC}\hskip .5cm (\rm{no\ sum}) .
\label{auto1}
\end{equation} 
For fermions, the requirement that
the coupling $igA_{M}^A{\overline \Psi_R}\gamma^{M}T^A{\Psi_R}$
is $Z_2$ invariant leads to the condition 
\begin{equation} \lambda_R T_R^{A} \lambda_R=\eta^A T_R^A, 
\label{automorphism2A} 
\end{equation} 
which can be simplified to
\begin{equation} [\lambda_R,T_R^a]=0\hskip 1cm \{\lambda_R,T_R^{\hat a}\}=0 
\label{auto2}. 
\end{equation} 

Some comments are now in order:

\begin{itemize}

\item
The presence of $i\gamma^5$ in (\ref{pfermion}) makes it that $\Psi_L$ and 
$\Psi_R$ have opposite parities. In particular this means that

\begin{itemize}
\item
Zero modes are {\it chiral}. Anomalies have then to be canceled as in
the Standard Model
\item
Non-zero modes are {\it non-chiral} and then anomaly free.
\end{itemize}

\item
The gauge group on the brane is $\mathcal{H}$ and only {\it even}
fields can couple to the brane. In particular for scalar fields only
$\partial^{\,2n}_5A_5^{\,\hat a}$ and $\partial^{\,2n+1}_5A_5^{\,a}$,
for $n=0, 1\dots$, can have non-zero couplings to the brane. We will
make use of this fact later on.

\end{itemize}

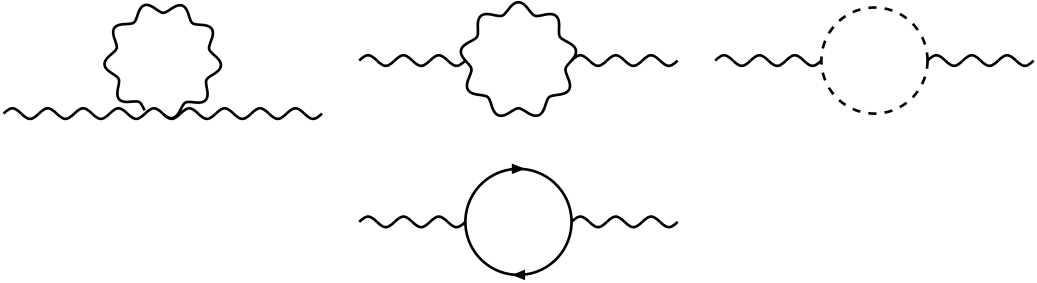
\begin{figure}[h]
\begin{center}
\begin{picture}(120,60)(0,0)
\Photon(0,0)(120,0)29
\PhotonArc(60,20)(20,-90,250)29 	
%\LongArrowArcn(60,20)(14,120,50)
%\footnotesize
%\Text(60,27)[c]{$q$}
%\Text(0,8)[l]{$m,A$}
%\Text(120,8)[r]{${m^\prime},A'$}
%\Text(40,30)[r]{$l,B$}
%\Text(80,30)[l]{$l',B'$}
%\LongArrow(45,-5)(75,-5)
%\Text(40,-6)[c]{$p$}
%\LongArrow(100,-5)(115,-5)
\end{picture}
\quad 
\begin{picture}(120,60)(0,0)
\PhotonArc(60,20)(20,0,360)29
\Photon(0,20)(40,20)23 
\Photon(80,20)(120,20)23	
%\LongArrowArcn(60,20)(14,120,50)
%\LongArrowArcn(60,20)(14,300,230)
%\footnotesize
%\Text(60,12)[c]{$r$}
%\Text(60,27)[c]{$q$}
%\Text(0,28)[l]{$m,A$}
%\Text(120,28)[r]{${m^\prime},A'$}
%\Text(41,0)[r]{$k,C$}
%\Text(41,40)[r]{$l,B$}
%\Text(79,0)[l]{$k',C'$}
%\Text(79,40)[l]{$l',B'$}
%\LongArrow(5,16)(20,16)
%\Text(12,10)[c]{$p$}
%\LongArrow(100,16)(115,16)
%\Text(108,10)[c]{$p$}
\end{picture}
\quad 
\begin{picture}(120,60)(0,0)
\DashCArc(60,20)(20,0,360)3
\Photon(0,20)(40,20)23 
\Photon(80,20)(120,20)23 
%\LongArrowArcn(60,20)(14,120,50)
%\LongArrowArcn(60,20)(14,300,230)
%\footnotesize
%\Text(60,12)[c]{$r$}
%\Text(60,27)[c]{$q$}
%\Text(0,28)[l]{$m,A$}
%\Text(120,28)[r]{${m^\prime},A'$}
%\Text(41,0)[r]{$k,C$}
%\Text(41,40)[r]{$l,B$}
%\Text(79,0)[l]{$k',C'$}
%\Text(79,40)[l]{$l',B'$}
%\LongArrow(5,16)(20,16)
%\Text(12,10)[c]{$p$}
%\LongArrow(100,16)(115,16)
%\Text(108,10)[c]{$p$}
\end{picture}
\begin{picture}(120,60)(0,0)
\ArrowArcn(60,20)(20,0,180)
\ArrowArcn(60,20)(20,180,360)
\Photon(0,20)(40,20)23 
\Photon(80,20)(120,20)23 
%\footnotesize
%\Text(60,6)[c]{$r$}
%\Text(60,33)[c]{$q$}
%\Text(0,28)[l]{$m,A$}
%\Text(120,28)[r]{${m^\prime},A'$}
%\Text(41,0)[r]{$k,C$}
%\Text(41,40)[r]{$l,B$}
%\Text(79,0)[l]{$k',C'$}
%\Text(79,40)[l]{$l',B'$}
%\LongArrow(5,16)(20,16)
%\Text(12,10)[c]{$p$}
%\LongArrow(100,16)(115,16)
%\Text(108,10)[c]{$p$}
\end{picture}
\end{center}
\caption{The one-loop diagrams contributing to mass and wave function 
renormalization}
\label{figura1}
\end{figure}

We have analyzed~\cite{vonGersdorff:2002as} bulk and brane
renormalizations effects from the one-loop diagrams in
Fig.~\ref{figura1}.
The wavy lines refer to 5D gauge bosons $A_M^A$, $M=\mu,5$ and
$A=a,\hat a$, the dotted lines to the FP-ghosts $c^A,\ \bar c^A$ and
the solid lines to the matter fermions $\Psi$ in the representation
{\bf R} of $\mathcal{G}$.

From the diagrams in Fig.~\ref{figura1} with external momentum
$p^\mu\neq 0$ there appear wave-function renormalization effects both
in the bulk $\sim z_0 F^A_{MN}F_A^{MN}$, and on the branes
\begin{equation}
\label{wfr}
\sim\left[\delta(x^5)+ \delta(x^5-\pi R)\right]\left\{z F^a_{\mu\nu}
F^{\mu\nu}_a+\widetilde{z} F^{\hat a}_{\mu 5}F^{\mu 5}_{\hat a}
\right\}
\end{equation}
This phenomenon was first analyzed in Ref.~\cite{Georgi:2000ks}. 

We have studied mass renormalization effects, i.e. corresponding to
zero external momentum $p^\mu=0$, for gauge fields $A^A_\mu$ and
scalar bosons $A_5^A$ from the corresponding diagrams in
Fig.~\ref{figura1}. In this note we will concentrate on the case of
external bosons $A_5^{\hat a}$ since they play the role of Higgs
bosons on the brane and can spontaneously break the gauge symmetry
$\mathcal{H}$ when they acquire a VEV.

\subsection{Bulk effects}

The most general terms that can appear in the bulk are given by the
Lagrangian
\begin{equation}
\mathcal{L}_{bulk}=A_5^{A}\Pi_{AB}[-\partial_5^2] A_5^{B}
\label{uno}
\end{equation}
In particular the expansion of the $\Pi_{\,A\,
B}[-\partial_5^2]$ operator around $p_5^2=0$ gives the squared mass
term as $\Pi_{AB}[0]=-M^2_{A}\delta_{\,A\,B}$, while the
other terms in the expansion correspond to higher dimensional
operators.

We obtain for the radiative mass in the bulk the expression
\begin{equation}
\label{masabulk}
M^2_{bulk}=\frac{3 g^2}{32\pi^4 R^2}\zeta(3)\left[ 3C_2(\mathcal{G})-4
\sum_R C_R\right]
\end{equation}
where $C_2(\mathcal{G})$ is the Casimir of the group $\mathcal{G}$,
$C_R$ the Dynkin index of the representation {\bf R} and the sum is
understood to be performed over all different fermions in the theory.
This result has been confirmed by the calculation of
Ref.~\cite{Cheng:2002iz}.

Here we can make two comments referring to the previous result:

\begin{itemize}
\item
The radiative mass (\ref{masabulk}) if {\it finite}. In fact it is
protected from quadratic divergences by the higher-dimensional gauge
invariance.

\item
Fermions induce negative squared masses. They can induce vacuum
expectation values for the $A_5^{\hat a}$ fields which in turn can
trigger spontaneous breaking of the gauge theory $\mathcal{H}$ on the
brane: the Hosotani breaking.

\end{itemize}

\subsection{Brane effects}

The general terms that can appear on the brane for the $A_5^{A}$
fields in the absence of space-time derivatives are of the form,
\begin{equation}
\mathcal{L}_{brane}=\left( \delta(x^5)+\delta(x^5-\pi R)\right)
A_5^{A}\widetilde{\Pi}_{A B}[i\!\stackrel{\leftarrow}{\partial_5},
i\!\stackrel{\rightarrow}{\partial_5}] A_5^{B}
\label{dos}
\end{equation}
and the arrows on the $\partial_5$ indicate the field whose derivative
is to be taken. Power expansion of $\Pi_{A
B}[i\!\stackrel{\leftarrow}{\partial_5},
i\!\stackrel{\rightarrow}{\partial_5}]$ around $p_5=0$ provides, as in
the case of bulk corrections, the different radiative corrections on
the brane. In particular the lowest order corrections can be written
as:
\begin{equation}
\label{bajos}
\mathcal{L}_{brane}=\left( \delta(x^5)+\delta(x^5-\pi R)\right)
\left\{\tilde z_1 (A_5^{\hat a})^2+\tilde z_2 A_5^{\hat
a}\partial_5^2A _5^{\hat a}+\tilde z_3 (\partial_5 A_5^{a})^2+\cdots
\right\}
\end{equation}

A one loop calculation yields~\cite{vonGersdorff:2002as}
\begin{equation}
\label{masabrane}
\widetilde\Pi_{A B}[\,p_5,p'_5]\equiv 0
\end{equation}
where the previous identity should be understood when the operator
$\widetilde\Pi_{A B}[\,p_5,p'_5]$ is bracketed as in
Eq.~(\ref{dos}). In particular this means that all the coefficients in
Eq.~(\ref{bajos}) vanish.

At first sight this result is a bit puzzling since even if the fields
$A_5^{\hat a}$, $\partial_5^2A _5^{\hat a}$, $\partial_5 A_5^{a}$ are
scalar fields in four dimensions, still the symmetries of the 5D
theory seem to protect these scalar fields on the brane from quadratic
divergences. In the next section we will identify the custodial
invariance on the brane.

\section{Restrictions on brane terms from higher dimensional gauge symmetries}

In this section we would like to clarify the implications of the
higher dimensional gauge symmetry ${\mathcal G}$ on possible brane
localized interactions. In the bulk, the local symmetry ${\mathcal G}$
is realized on the gauge fields $A_M^A$ as
\begin{equation}
\delta_{\mathcal G} A^A_M=\partial_M \xi^A + f^{ABC}\xi^B A^C_M
\label{bulkgauge}
\end{equation}
After orbifolding the gauge symmetry is still intact provided the
parity assignments for the fields $A_M^A$ are chosen to satisfy the
automorphism condition (\ref{automorphism1}) consistent with the
orbifold structure, as well as a similar condition for the gauge
parameters
\begin{equation}
\label{autoxi}
\xi^A(x^\mu,-x^5)=\Lambda^{AB}\xi^B(x^\mu,x^5).
\end{equation}

On the branes ${\mathcal G}$ is broken down to the subgroup ${\mathcal
H}$. The nonzero fields are the gauge bosons of ${\mathcal H}$,
$A_\mu^a(0)$, and the scalars $A_5^{\hat a}(0)$. Evaluating
Eq.~(\ref{bulkgauge}) at $x_5=0$ one gets for transformations
generated by $\xi^{a}$
\begin{eqnarray}
\delta_{\mathcal H} A^a_\mu&=&\partial_\mu \xi^a +
f^{abc}\xi^b A^c_\mu,\label{branegauge1}\\ 
\delta_{\mathcal H} A^{\hat
a}_5&=&f^{\hat{a}b\hat{c}}\xi^b A^{\hat c}_5.
\label{branegauge2}
\end{eqnarray}
All fields and parameters are understood to be evaluated at $x_5=0$
and only depend on $x^\mu$. The other terms present in
Eq.~(\ref{bulkgauge}) can be seen to vanish by parity by using that
only the structure constants with zero or two hatted indices are
nonzero, i.e. $f^{\hat a bc}=f^{\hat a\hat b\hat c}=0$.
Eq.~(\ref{branegauge1}) just means that $A_\mu^a(0)$ transforms as a
gauge boson for the brane gauge symmetry ${\mathcal H}$, while
Eq.~(\ref{branegauge2}) shows that the scalars transform in a
representation given by
\begin{equation}
\left( T^b \right)_{\hat a\hat c}\equiv if^{\hat{a}b\hat{c}}.
\label{representation}
\end{equation}
It is easy to show that the $T^b$ indeed form a (not necessarily
irreducible) representation of $\mathcal H$.  All brane interactions
must respect the symmetry ${\mathcal H}$ acting as in
Eqs.~(\ref{branegauge1}) and (\ref{branegauge2}).

However, there is a further remnant of ${\mathcal G}$ on the
branes~\footnote{We thank R.~Sundrum for bringing this to our minds.}:
Doing the same reduction of Eq.~(\ref{bulkgauge}) but this time for
the odd parameters $\xi^{\hat a}$ one gets
\begin{eqnarray}
\delta_{\mathcal K} A^a_\mu&=&0,\label{braneremnant1}\\ 
\delta_{\mathcal K} A^{\hat a}_5&=&\partial_5\xi^{\hat a}.
\label{braneremnant2}
\end{eqnarray}
Eq.~(\ref{braneremnant2}) is a local (i.e.~$x^\mu$-dependent)
${\mathbb R}^{d_{\mathcal K}}$ shift symmetry (and it is worthwhile to
notice that it takes the same form for finite $\xi$ as well).  Its most
important implication is that it forbids mass terms for the scalars on
the branes as they appear for instance in Eq.~(\ref{bajos}) since
\begin{equation} 
\delta_{\mathcal K}\left( A_5^{\hat a}M_{\hat a\hat c}A_5^{\hat c}
\delta(x_5)\right) =2 A_5^{\hat a}M_{\hat a\hat c}\partial_5\xi^{\hat
c}\delta(x_5)\neq 0.  
\end{equation}
In principle, one can derive infinitely many~\footnote{It is rather
counterintuitive that the single gauge transformation
Eq.~(\ref{bulkgauge}) implies infinitely many transformations of the
fields on the branes. This is due to the dependence $\xi=\xi(x_5)$
which translates to infinitely many independent transformation
parameters $\left.\partial_5^n\xi^{\hat a}\right|_{x_5=0}$.}
constraints from the original gauge symmetry by taking powers of
$\partial_5$ of Eq.~(\ref{bulkgauge}) and evaluating at $x_5=0$. This
yields for example
\begin{equation}
\delta_{\mathcal K}\left( \partial_5A_\mu^{\hat a} \right)=
\left( {\mathcal D}_\mu \right)^{\hat a \hat c} \left(\partial_5 \xi^{\hat c} \right),
\end{equation}
where ${\mathcal D}_\mu=\partial_\mu+iA_\mu^bT^b$ is the ${\mathcal
H}$-covariant derivative in the representation
Eq.~(\ref{representation}).  From this one sees that
\begin{equation}
\left( {\mathcal D}_\mu \right)^{\hat a \hat c}A_5^{\hat c}-\partial_5A_\mu^{\hat a} 
\label{fieldstrength}
\end{equation}
has vanishing variation $\delta_{\mathcal K}$ which is not surprising
since Eq.~(\ref{fieldstrength}) is just the field strength $$F_{\mu
5}^{\hat a}(x_5=0)$$ which appears from the wave function
renormalization on the brane, Eq.~(\ref{wfr}). Along these lines one could
obtain all brane symmetries induced by the original gauge symmetry
Eq.~(\ref{bulkgauge}) and so restrict the possible localized terms.

\section{Conclusions}

In this note we have explored a non-supersymmetric alternative to the
solution of the hierarchy problem (here understood as the appearance of
quadratic divergences for the Higgs boson mass). This alternative is
realized if electroweak symmetry breaking is triggered by the vacuum
expectation value of the extra dimensional component of a higher
dimensional gauge boson: the Hosotani mechanism.

A higher dimensional theory is non-chiral and to achieve
four-dimensional chirality it requires an orbifold
compactification. For this mechanism to hold the higher dimensional
gauge group $\mathcal{G}$ must be broken by orbifolding to the
Standard Model gauge group in such a way that the corresponding Higgs
bosons have the Standard Model quantum numbers and interactions.

The Higgs mass is zero at tree-level. It is protected in the bulk from
quadratic divergences by the higher-dimensional gauge invariance. It
can however get by radiative corrections a tachyonic mass $\sim 1/R$
that can break spontaneously the Standard Model by the Hosotani
mechanism.

For the Higgs mass we have found no one-loop quadratic divergences
localized on the four-dimensional branes in spite of the fact that
scalar Higgs boson masses are not protected by any four-dimensional
symmetry on the brane. There is however a remnant of the
higher-dimensional gauge invariance that forbids such mass terms. This
custodial symmetry should be effective at any order in perturbation
theory and so we do not expect any mass counterterms at higher
loops. Of course this does not mean that there are not higher-loop
divergences since they will be implemented by wave-function
renormalization on the brane. This divergence should however be
absorbed into the running of the gauge coupling.

Let us finally mention that this mechanism is suitable to low-scale
string models and points towards the unification of electroweak theory
into an $SU(3)$ group as the simplest possibility with
$\sin^2\theta_W\simeq 0.25$.

\section*{Acknowledgments}
One of us (MQ) would like to thank the Physics and Astronomy
Department of the Johns Hopkins University, where part of this work
has been done, for the hospitality extended to him as Bearden Visiting
Professor.  This work is supported in part by EU under RTN contracts
HPRN-CT-2000-00148 and HPRN-CT-2000-00152 and in part by CICYT, Spain,
under contract FPA2001-1806. The work of GG was supported by the DAAD.

\end{document}